# Insight into the charge density wave gap from contrast inversion in topographic STM images

M. Spera,[1,#] A. Scarfato,[1,#] Á. Pásztor,[1] E. Giannini,[1] D.R. Bowler,[2] and Ch. Renner[1,*]

[1]Department of Quantum Matter Physics, University of Geneva,

24 Quai Ernest-Ansermet, CH-1211 Geneva 4, Switzerland

[2]London Centre for Nanotechnology and Department of Physics and Astronomy,

University College London, London WC1E 6BT, United Kingdom

## Abstract

Charge density waves (CDWs) are understood in great details in one dimension, but they remain largely enigmatic in two dimensional systems. In particular, numerous aspects of the associated energy gap and the formation mechanism are not fully understood. Two long standing riddles are the amplitude and position of the CDW gap with respect to the Fermi level ($E_F$) and the frequent absence of CDW contrast inversion (CI) between opposite bias scanning tunneling microscopy (STM) images. Here, we find compelling evidence that these two issues are intimately related. Combining density functional theory and STM to analyse the CDW pattern and modulation amplitude in 1$T$-TiSe$_2$, we find that CI takes place at an unexpected negative sample bias because the CDW gap opens away from $E_F$, deep inside the valence band. This bias becomes increasingly negative as the CDW gap shifts to higher binding energy with electron doping. This study shows the importance of CI in STM images to identify periodic modulations with a CDW and to gain valuable insight into the CDW gap, whose measurement is notoriously controversial.

## Introduction

The charge density wave (CDW) ground state is an atomic length scale periodic modulation, combining lattice and charge degrees of freedom [1]. The precise mechanism driving this phase transition remains largely unknown. Fermi surface nesting, electron-electron or electron-phonon interactions, and coupling of electrons to other degrees of freedom in the host crystal are among the main mechanisms discussed over the years [2, 3].

Below the CDW phase transition temperature, atoms rearrange into periodic lattice distortions. Concomitantly, charge is redistributed in real space to form alternating regions of charge accumulation and charge depletion. In the classic Peierls mechanism, mostly states in the vicinity of the Fermi level ($E_F$) are involved in the CDW formation and a gap opens at $E_F$.

Scanning tunneling microscopy (STM), owing to its high spatial topographic resolution, is an ideal probe to characterize the real space charge ordering. In the Peierls scenario, constant current STM images of the CDW depend on sample bias voltage ($V_b$) polarity: negative bias will show enhanced intensity over charge accumulation regions, whereas images of the same area at positive bias will show enhanced intensity over charge depleted regions. This is known as contrast inversion (CI) of the CDW STM pattern.

Contrast inversion is often considered a hallmark of the CDW contribution to the STM topographic signal. However, clear CI between opposite polarity STM images has only been reported in very rare cases [4, 5], including high temperature superconductors [6] and transition metal dichalcogenides (TMDs) [7]. In a thorough theoretical analysis for 2$H$-NbSe$_2$, Sacks et al. [8] conclude that CI does not take place in this material due to band structure effects. They further contend that CI is in general not expected for two-dimensional CDW systems. In a more recent STM study of TaS$_2$, TaSe$_2$ and NbSe$_2$, Dai et al. [9] conclude that strong lattice distortions completely mask possible CI arising from electronic contributions to the CDW amplitude in the topography.

In addition to the real space reconstruction introduced above, the CDW ground state is also characterized by a gap in the electronic density of states (DOS). For strong coupling CDW materials the gap does in general not open for all momenta or necessarily at the Fermi level [10, 11]. In particular for TiSe$_2$, the Ti $3d_{z^2}$ and the Se $4p_z$ bands only marginally participate in the CDW reconstruction [12-14]. Moreover, the electronic nature of this compound is still a matter of debate. For instance, recent angle-resolved photoemission spectroscopy (ARPES) studies find either semiconducting [13] or semimetallic [14] behaviour with the bands affected differently upon the CDW transition.

Since the CDW gap opens only in a small portion of the band structure and scanning tunneling spectroscopy (STS) is a momentum averaging technique, STS spectra lack distinct spectroscopic features allowing a sure identification of a momentum dependent CDW gap. Moreover, clear disentanglement of CDW related reduction of the DOS from band structure features are further complicated in TiSe$_2$ due to its debated electronic nature. Therefore, determining the amplitude and position of the CDW gap with respect to $E_F$ by STS is problematic. Here, instead of spectroscopic data, we focus on bias dependent topographic images, which we demonstrate to provide novel insight into the CDW gap and contrast inversion. Supported by DFT simulations and a simple one-dimensional model, topographic images show that the absence of CI in opposite polarity STM images is a direct consequence of the CDW gap not opening at $E_F$. When present, CI informs about the CDW gap amplitude, which is often significantly larger than expected from the phase transition temperature. Finally, detailed analysis of CI provides direct clues about the inadequacy of Fermi surface nesting as the main mechanism driving the CDW phase transition.

**Results and Discussion**

Simple visual inspection of the constant current STM images presented in Fig. 1 (top row) shows that CDW contrast inversion does not happen between the opposite polarities data, but between the frames acquired at $-300$ mV and $-100$ mV. These images acquired at different bias voltages were aligned with atomic scale precision using well-identified single atom oxygen and titanium defects resolved in large-scale images [15]. In contradiction to these experimental findings, detailed DFT calculations for undoped pristine 1$T$-TiSe$_2$ predict CDW contrast inversion between $-100$ mV and $+100$ mV (Fig. 1, bottom row), as expected in the classic Peierls model.

DFT does reproduce the experimental observation when doping electrons into the unit-cell (Fig. 1, middle row). The energy where CI takes place in this case depends on carrier concentration, shifting to higher binding energy with increasing electron content. Interestingly, there is a practical limit as to how far CI can be observed at negative bias (see Supplemental Material Sec. II for extended DFT data [16]). Indeed, for high enough doping, CDW contrast disappears altogether before the negative bias voltage where contrast inversion would actually appear.

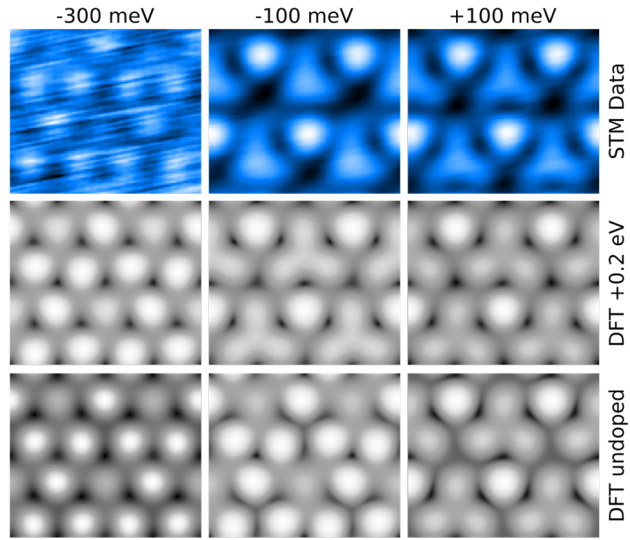

**Figure 1.** Charge order contrast inversion revealed by STM on 1$T$-TiSe$_2$. Top row: $1.4 \times 1.2$ nm$^2$ bias dependent STM micrographs of the same area showing contrast inversion below the Fermi energy between the two images recorded at $V_b = -300$ mV and $V_b = -100$ mV. Set parameters of STM data are, from left to right, $V_b = -300$ mV and $I_t = 600$ pA, $V_b = -100$ mV and $I_t = 100$ pA, $V_b = 100$ mV and $I_t = 200$ pA. The Z-range of the STM data from left to right is 4.3 pm, 37.2 pm and 36.1 pm. Middle and bottom rows show the DFT simulations of the expected STM topographic contrast as a function of bias in electron doped and in pristine 1$T$-TiSe$_2$, respectively.

The STM data and DFT simulations of Fig. 1 reveal a remarkable doping dependence of the energy where CI occurs. The disagreement between experiment and DFT for undoped TiSe$_2$ is likely to arise from two different sources: first, the well-known problem with DFT underestimating band gaps means that the fine details of the TiSe$_2$ gap will not be perfectly accurate; second, the well-known self-doping of TiSe$_2$ with excess Ti in experiments, which will shift the Fermi level. Bias dependent STM images of the CDW near single atom defects also suggest a doping dependence of the CI. Most of the area imaged in Fig. 2 does not show any CI between −100 mV and +100 mV, as already pointed out in Fig. 1. However, there is contrast inversion between these two biases in the right-hand side region of Fig. 2e and 2f, between the defects marked A and B. This region with inverted contrast is expanding with increasing negative sample bias to encompass most of the field of view at −300 mV (Fig. 2a–d).

Comparing with DFT simulations, the absence of CI at opposite biases in the central regions of Fig. 2, away from the defects, is consistent with a globally electron-doped system. The contrast around the defects indicates that the doping is locally modified: while defects C and D behave markedly as electron donors, the presence of CI near zero bias in the region between defects A and B suggests that they have a light hole-doping character, such that their combined action turns the area towards neutrality. The result is an increasing electron doping gradient from defects A and B to defects C and D, with the strongest electron doping character around defect D [17]. The role of defects and bias dependent STM imaging has been discussed previously in 2$H$-NbSe$_2$ [18]. However, the focus of that study was on the ability of defects to stabilize the CDW phase near the transition temperature without addressing CI and local doping effects. Note that while we observe a marked spatial dependence of the CI due to local doping, the corresponding shift of the STS spectra is very small (see Supplemental Material Fig. S2 for additional STS data [16]).

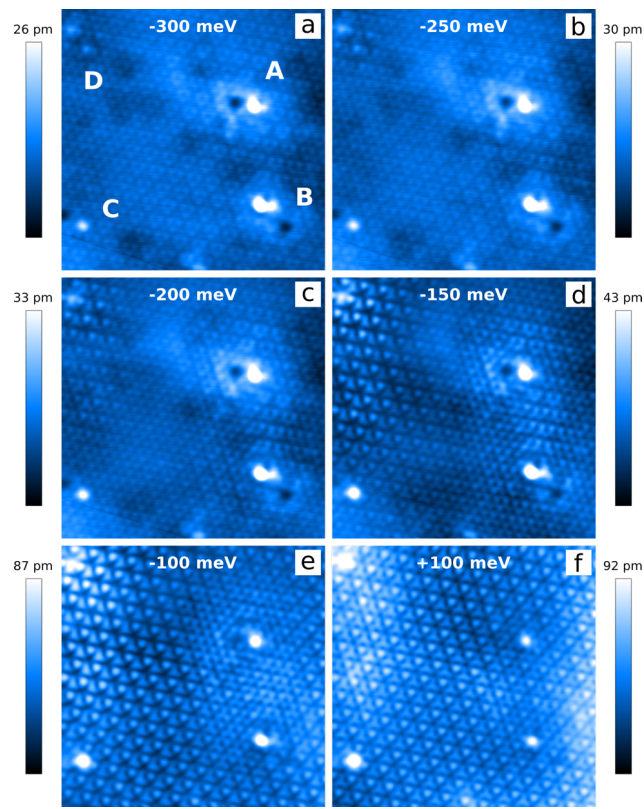

**Figure 2.** Doping dependence of the CDW contrast inversion. $10 \times 10$ nm$^2$ STM topography of 1$T$-TiSe$_2$ at (a) −300 mV, (b) −250 mV, (c) −200 mV, (d) −150 mV, (e) −100 mV and (f) +100 mV sample bias. CI happens at an increasingly higher binding energy near defects A to D, indicating their different doping nature (see text and Fig. S3 in Supplemental Material, Sec IV [16]). Set currents are (a) 600 pA, (b) 300 pA, (c, d and f) 200 pA, (e) 100 pA.

As mentioned in the introduction, there are very few topographic STM studies reporting contrast inversion for any CDW material in the literature [4, 9, 19]. In most cases, the focus has been on comparing images taken at opposite polarities, as would be expected in a classic Peierls transition. Here, we find that images taken at low opposite biases are indeed very similar, while clear CI is observed between images measured at selected negative biases. In the following, we demonstrate that this observation is a direct consequence of a CDW gap opening below $E_F$ (corresponding to $V_b = 0$ V) and shifting to higher binding energy with increasing electron doping.

First, we demonstrate that the observed contrast inversion in the STM micrographs is indeed a CDW feature. To this end, we consider two real space images of the CDW taken at two different biases below $E_F$. Fig. 3a and 3b are magnifications from Fig. 2b and 2e of the uniform region in between the defects A-D, where we decided to keep defect A, visible in the top right corner of the cropped images, as a reference point. Fig. 3a was acquired at a large negative bias, where the periodic pattern reflects the charge accumulation associated with the CDW reconstruction in TiSe$_2$. Fig. 3b was acquired at a smaller negative bias closer to the Fermi level, and shows the corresponding periodic pattern of charge depletion.

If the differences between the two images in Fig. 3a and 3b are due to the CDW, they should be related through an inversion of the periodic charge order reconstruction pattern. To verify this, we numerically invert the contrast of the CDW signal of the charge depletion image in Fig. 3b in the following way. First, we apply Fourier filtering to separate the signal corresponding to the CDW from the rest of the image. We then invert this CDW image and recombine the result with the rest of the original micrograph (see Supplemental Material Sec. V for more details [16]). The resulting image shown in Fig. 3c is in excellent agreement with the experimental charge accumulation image in Fig. 3a. Alternatively, if we invert not only the CDW component, but the full image, including the atomic lattice contrast in Fig. 3b, we obtain a completely different topographic pattern unable to reproduce the experimental one (Fig. 3d). The contrast difference between the two STM topographies in Fig. 3a and 3b is thus definitely due to the CDW contrast inversion.

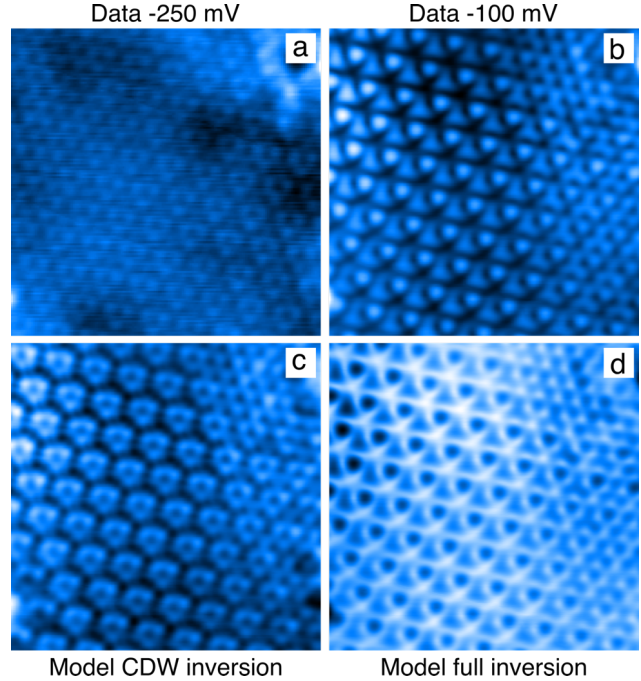

**Figure 3.** Charge order origin of the contrast inversion observed in bias dependent STM images of 1$T$-TiSe$_2$. (a) Negative sample bias image at $-250$ mV showing the real space distribution of charge accumulation (set current is 300 pA, z-range = 33 pm). (b) Negative sample bias image at $-100$ mV showing the real space distribution of charge depletion (set current is 100 pA, z-range = 66 pm). (c) Contrast inversion applied only to the CDW component of the image in (b). (d) Contrast inversion applied to all components of the image in (b). Image size is $6 \times 6$ nm$^2$.

An interesting aspect of the STM images in Fig. 2e-f is that we not only observe the same CDW pattern at $\pm 100$ mV opposite polarities in some regions, they also show the same CDW modulation amplitude defined as the peak-to-peak amplitude of the local CDW topographic signal (see Section VI in the Supplementary for the technical details of the analysis [16]). Such close correspondence (both in phase and amplitude) between opposite polarity topography is not possible if the CDW gap opens across the Fermi level as we show in the simple one-dimensional model illustrated in Fig. 4.

Let us consider a one-dimensional (1D) BCS-like local density of states (LDOS) [9], with a partial gap centered at $E_F$ (Fig. 4c) and a harmonic spatial modulation where the states above and below the gap midpoint are spatially 180° degrees out-of-phase (Fig. 4a). A constant current STM image amounts to integrating over all states from $E_F$ up to the imaging bias at each sampling point. Similarly, we reconstruct a bias dependent topography by integrating our model LDOS (Fig. 4a) between $E_F$ and the imaging bias at each position, including a finite thermal smearing (Fig. 4b). When the CDW gap is centered on $E_F$ (Fig. 4a), the integration at positive (negative) bias runs over primarily depleted (accumulated) states. In this case, CDW CI (equivalent to a $\pi$-shift for a harmonic density profile) is expected between opposite polarity images. This can be seen explicitly as a $\pi$-phase shift at $E_F$ (Fig. 4c) in the phase as a function of imaging bias of the calculated sinusoidal CDW signal in Fig. 4b. This situation certainly does not describe our experimental data. Indeed, in this case, one does not expect to image the same CDW pattern at positive and negative sample bias, contrary to our observations reproduced in the Figs. 1 and 2.

A very different bias dependence of the phase appears when shifting the CDW gap below the Fermi level, as illustrated in Fig. 4f. In this case, positive bias STM imaging still involves integration over primarily depleted states as in the absence of a shift. However, imaging up to a finite negative bias corresponding to the middle of the CDW gap ($-V_{mid}$) will still reflect depleted states (Fig. 4d), in contrast to the unshifted case. Hence, the CDW pattern corresponds to depleted charge regions when imaging at biases between $-V_{mid}$ and a finite positive bias. No CI is expected between opposite polarity images in the range $\pm V_{mid}$ (Fig. 4e, solid and dotted red lines), in agreement with experiment.

Reducing the bias voltage below $-V_{mid}$, both charge depletion and charge accumulation in the local density of states contribute to the tunneling current. Within the harmonic and symmetric charge redistribution model considered here, the CDW contrast is then progressively reduced to ultimately disappear at a compensation bias voltage $-V_{comp} = -2\,V_{mid}$ (Fig. 4e, dotted black line). One will have to set the bias voltage below $-V_{comp}$ to reveal the CDW pattern corresponding to charge accumulation. Consequently, contrast inversion does not happen at $E_F$, but at a negative bias $-V_{comp}$ corresponding to

$$\int_{-V_{comp}}^{-V_{mid}} LDOS \cdot dE = \int_{-V_{mid}}^{0} LDOS \cdot dE.$$

Note that $-V_{comp}$ can be larger than the maximum negative bias where CDW contrast is still achievable, in which case contrast inversion cannot be observed. This is the case, for example, in Cu intercalated TiSe$_2$ [17]. The actual values of $-V_{mid}$ and $V_{comp}$ will depend on the detailed material band structure.

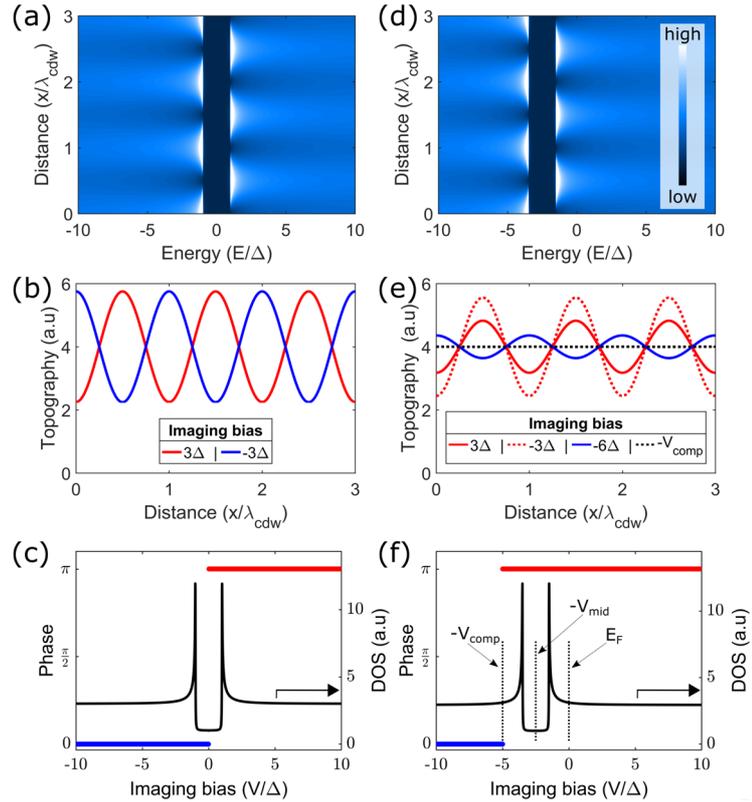

**Figure 4.** One-dimensional model description of the CDW contribution to the STM topography. (a) Spatial and energy dependent CDW LDOS with the CDW gap centered at $E_F$. (b) Corresponding topography traces with inverted contrast between opposite polarity traces, and (c) their bias dependent phases: contrast inversion happens at $E_F$. (d) Spatial and energy dependent CDW LDOS with the CDW gap shifted below $E_F$. (e) Corresponding topography traces at selected bias voltages and (f) their bias dependent phases: contrast inversion happens below $E_F$. The solid black line in panel (c) and (f) illustrate the BCS-like model DOS used in the modeling.

In case the gap is centered at $E_F$, the CDW contrast is zero for all biases within the gap, or vanishingly small when including thermal smearing effects (see Fig. S5 in Supplemental Material [16]). The situation changes completely when the gap is shifted below the Fermi level (Fig. 4f). As shown above, the bias range where STM imaging involves integration over primarily depleted states is no longer confined to positive biases, but extends to finite negative biases. At these negative biases, STM imaging will show the same CDW pattern as at positive bias. Remarkably, for a large enough shift of the gap, we do not only expect the same pattern but also the same CDW modulation amplitude at selected opposite polarity biases as shown in Fig. S5b in the Supplemental Material (see Sec. VI in Supplemental Material for a detailed comparison of the data with the model) [16]. Another striking feature is that the CDW amplitude will remain finite for biases inside the CDW gap when the latter is shifted below $E_F$. This is in contrast to the absence of CDW contrast for biases below the gap when it is opening at $E_F$. Whether the CDW gap is at $E_F$ or below, the CDW contrast will vanish for large enough imaging biases when the weight of the reconstructed states relative to the unreconstructed states becomes too small to resolve.

The simple 1D model introduced in Fig. 4 reproduces all the topographic CDW features observed by STM on TiSe$_2$. Considering the bias dependent CDW modulation amplitude in a region away from any defect, we can make a rough estimate of the gap amplitude and position below $E_F$ (see Supplemental Material Fig. S5 [16]). For a region near the middle of Fig. 2, we find a good correspondence between the data and the 1D model for a CDW gap width of $\Delta = 70$ meV and a shift of $\cong 1.3\,\Delta$. This result is perfectly consistent with the rigid band shift moving the CDW gap to higher binding energy as a function of electron doping observed by ARPES [20, 21] and STM [17]. In our experiments, we observe local variations of the doping and associated band shift near atomic impurities, which provide a unique opportunity to verify our model predictions in a single experiment. Depending on sample bias and distance from these atomic dopants, we observe all three contrast configurations discussed in Fig. 4e. STM images reveal CDW patterns corresponding to charge accumulation regions, to charge depletion regions and regions without contrast corresponding to imaging at $-V_{comp}$ (Fig. 2). Note that finding CI at a finite negative sample bias in pristine TiSe$_2$ is consistent with electron doping due to the unavoidable Ti self-doping.

## Conclusions

The comprehensive STM study presented here provides an alternative experimental insight into the amplitude and position of the CDW gap with respect to $E_F$, which both are still controversial [22-27]. Analyzing the bias dependence of the CDW modulation amplitude and phase, we find compelling evidence that the gap is not pinned to the Fermi level. This directly explains the absence of CDW contrast inversion between opposite polarity STM images as would be expected in the classic Peierls description. Such insight may prove instrumental in associating unknown periodic structures in STM images with a CDW. We demonstrate that CI in TiSe$_2$ can take place at negative bias voltages significantly away from the Fermi level, with a remarkable dependence on the local doping. We find this dependence is a direct consequence of the bands and the CDW gap shifting to higher binding energy upon electron doping. The simple model we propose further explains the absence of CI in Cu doped TiSe$_2$ as due to the CDW gap shifting significantly below the Fermi level [17]. The doping dependent CI we observe by STM poses explicit constraints on any model description of the CDW phase transition in TiSe$_2$. It suggests in particular that the CDW formation involves primarily electronic states away from the Fermi level, which implies that the transition cannot be driven by a particular topology of the Fermi surface. The system gains energy through the momentum dependent electron-phonon and electron-electron interactions, emphasizing the strongly correlated nature of electrons in the CDW phase of TiSe$_2$.

## Acknowledgements

This project was supported by the Swiss national science foundation through Div.II (grant 162517). We acknowledge stimulating discussions with J. van Wezel, B. Hildebrand, T. Jaouen, T. Gazdic, Ch. Berthod, and J. Lorenzana. We thank C. Barreteau for her help with characterizing the single crystals via transport measurements, and G. Manfrini and A. Guipet for their skillful technical assistance.

## Author Contributions

C.R. designed the experiment. A.S. and M.S. performed the STM measurements. M.S., A.S. and Á.P. performed data analysis. D.B. performed the DFT simulations. Á.P. conceived the one-dimensional model. E.G. synthesized the bulk crystals. C.R., Á.P. and A.S. wrote the paper. All authors contributed to the scientific discussions and manuscript revisions.

\# These authors made equal contributions to the work

\* To whom correspondence should be addressed

# Supplemental Material for

### Insight into the charge density wave gap from contrast inversion in topographic STM images

M. Spera,[1,#] A. Scarfato,[1,#] Á. Pásztor,[1] E. Giannini,[1] D.R. Bowler,[2] and Ch. Renner[1,*]

[1]Department of Quantum Matter Physics, University of Geneva,
24 Quai Ernest-Ansermet, CH-1211 Geneva 4, Switzerland

[2]London Centre for Nanotechnology and Department of Physics and Astronomy,
University College London, London WC1E 6BT, United Kingdom

## I. Crystal growth and STM measurements

Single crystals of 1$T$-TiSe$_2$ were grown via iodine vapor transport. A stoichiometric mixture of titanium and selenium was sealed in a quartz ampoule under vacuum. Crystals were subsequently grown at T=590°C for 25 days, to minimize the unavoidable Ti self-doping [1]. We performed scanning tunneling microscopy and spectroscopy using a SPECS Joule-Thomson STM with base temperature of 1.2 K and base pressure below 1·10$^{-10}$ mbar. Tips were made of mechanically cut PtIr and electrochemically etched tungsten wires, with no noticeable differences in the data. Each tip was conditioned in-situ on single crystal Ag(111) or Au(111) surfaces. The TiSe$_2$ single crystals were cleaved in ultra-high vacuum at room temperature. All the STM data were acquired at 5 K, and negative bias corresponds to probing states below the Fermi level of the sample.

## II. DFT modelling

DFT modeling was performed with the plane wave pseudopotential code VASP [2, 3], version 5.3.5. Projector-augmented waves [4] in a 7.01×7.01 A$^2$ rhombohedral unit cell were used with the Perdew-Burke-Ernzerhof [5] exchange correlation functional and plane wave cutoff of 212 eV. The 1$T$-TiSe$_2$ surface was modeled with four layers. A Monkhorst-Pack mesh with 9×9×1 k-points was used to sample the Brillouin zone of the cell. The parameters gave an energy convergence better than 0.01 eV. During structural relaxations, a tolerance of 0.03 eV/Å was applied. STM images were generated following the Tersoff-Hamann [6] approach in which the I(V) characteristic measured by STM is proportional to the integrated local density of states (LDOS) of the surface using the BSKAN code [7]. The effect of doping was simulated by adding a fraction of an electron per unit cell (up to 0.4) with an accompanying uniform positive background.

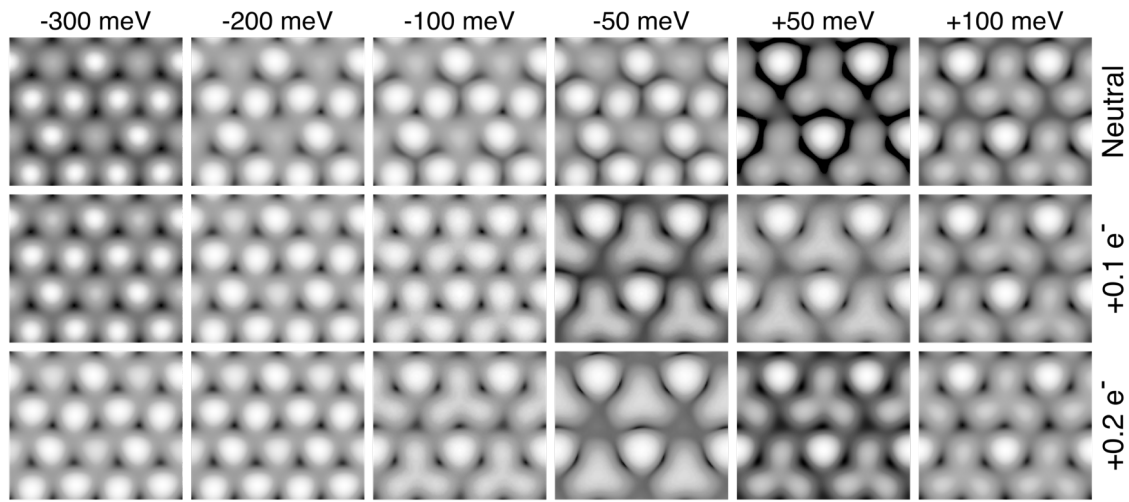

**Figure S1.** DFT simulations as a function of bias and electron doping. Without electron doping (first row) the contrast inversion happens across zero bias. Adding fractions of electrons per unit-cell, we find that the contrast distinctive of positive bias images of undoped specimen is now observed at negative biases, at −50 meV for +0.1 $e^−$ doping and at −100 meV for +0.2 $e^−$ doping per unit cell.

## III. STS spectroscopy

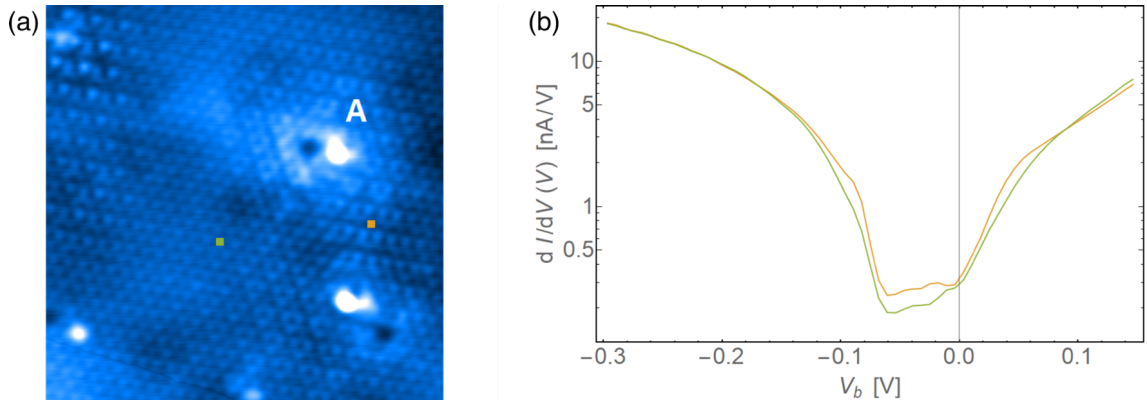

**Figure S2.** (a) 10×10 nm$^2$ STM topography of 1$T$-TiSe$_2$ - same image as in Fig. 2c of the main text. (b) Average of several d$I$/d$V$($V$) spectra measured over the areas marked with corresponding color dots in panel a. The areas correspond to the same location in the CDW unit cell, i.e., they are connected by a lattice vector. While we find a surprisingly large shift in CI bias between the two regions, the corresponding d$I$/d$V$ spectra reveal only minute spatial variations. See also ref. [8] for a detailed STS investigation of Cu doped TiSe$_2$.

## IV. Spatial (doping) dependence of CI as a function of bias

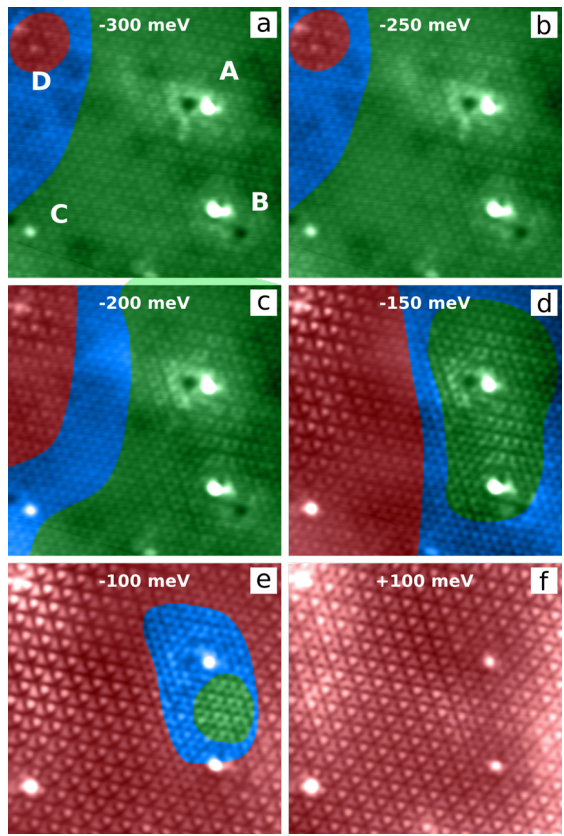

**Figure S3.** Reproduction of Fig. 2 with added color coding highlighting the shrinking (expansion) of the regions colored in green (red) with inverted (direct) contrast, while sweeping the bias from energies below to above the CDW gap. The blue color highlights regions with faint or no CDW contrast. The effect of doping is demonstrated by the persistence of the direct contrast in a region surrounding defect D down to −300 mV and by the inverted contrast in-between defects A and B at −100 mV.

## V. Contrast inversion by Fourier filtering

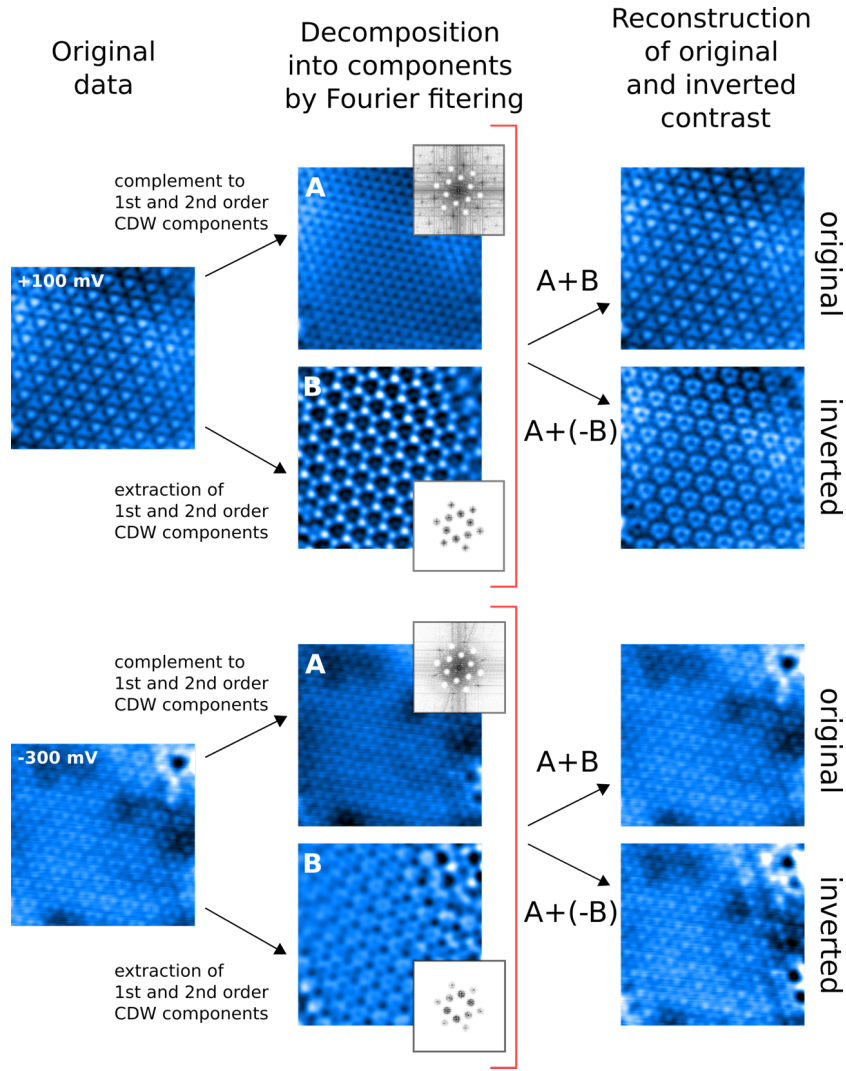

**Figure S4.** Illustration of the Fourier filtering technique applied to STM topographies acquired at biases above and below the CDW gap. The original data are first decomposed by Fourier filtering into two images containing (B) the 1st and 2nd order CDW components and (A) the complement to B. The filtered Fourier transforms used are reproduced in the insets with inverted gray color scale: for A, we employ a circular band stop filter centered on the most unambiguous CDW components (the 1st order CDW Bragg peaks and their sums); for B, we use a band pass filter centered on the same peaks. For each bias, adding A+B trivially reconstructs the original data, while summing A and -B, i.e. the filtered image with the CDW components inverted, reproduces the STM contrast observed at biases on the opposite side of the CDW gap. The atomic lattice A is the same for both biases, while the CDW reconstruction is different and shifted by one atomic position between the two bias voltages, corresponding to a $\pi$ phase shift.

## VI. Calculation of CDW amplitude in STM

We simulated the CDW spatial dependent LDOS by using a mean field model as derived in [9, 10]

$$\rho(x, E) = \rho_0 + \text{sgn}(E)\left[1 - \frac{\Delta}{E + i\Gamma}\cos\left(\frac{2\pi x}{\lambda_{CDW}}\right)\right]\frac{E + i\Gamma}{\sqrt{(E + i\Gamma)^2 - \Delta^2}} \quad (1)$$

where $E$ is the quasiparticle energy with respect to $E_F$, $\Gamma$ is a broadening term, $\Delta$ is the CDW gap amplitude, $\lambda_{CDW}$ is the CDW wavelength and $\rho_0$ accounts for the fact that the gap does not open for all momenta. The gap shift $\epsilon$ was introduced by replacing $E \rightarrow E - \epsilon$.

For the calculation of the STM CDW amplitude we start from the 1D Bardeen model [11]:

$$I(x, z, V) = \frac{4\pi e}{\hbar}\int_{-\infty}^{+\infty}\rho_t(E - eV)\rho_s(x, E)\,M(z, V, E)[f(E - eV, T) - f(E, T)]dE \quad (2)$$

where $M(z, V, E)$ is the transmission factor for a one-dimensional trapezoidal barrier, $\rho_s$ and $\rho_t$ are the sample and tip DOS respectively, $z$ is the tip-sample distance, and $V$ is the bias voltage.

Thermal smearing included via the Fermi function and the broadening term $\Gamma$ in the sample DOS are not equivalent. However, in the energy range considered here, their effect is not qualitatively different. Including the thermal smearing in the broadening term $\Gamma$ is numerically easier to handle and the only slight difference is in the smoothing of the peaks. We thus make the choice to include a small broadening by setting $\Gamma = 0.001$ and taking equation 2 in the limit $T \rightarrow 0$. By further assuming a constant tip DOS $\rho_t = 1$, and replacing the explicit expression of the transmission factor, the integral equation for the current simplifies as:

$$I(x, z, V) = \frac{4\pi e}{\hbar}\int_0^{eV}\rho(x, E)\,e^{-\kappa(E, V)z}dE \quad (3)$$

where $\kappa(E, V) = \frac{2\sqrt{2m}}{\hbar}\sqrt{\phi + \frac{eV}{2} - E}$ and $\phi$ is the effective work function.

To find the CDW amplitude we solved numerically the equation $I(x, z, V) = I_{set}$ for $z$ as a function of position and bias to obtain a map $z(x, V)$. Then the CDW amplitude as a function of bias voltage has been extracted as $A_{CDW}(V) = \max_x z(x, V) - \min_x z(x, V)$. We calculated $A_{CDW}(V)$ for several CDW gap sizes in the range 20 meV to 90 meV and with relative gap shifts $\epsilon/\Delta$ from 0 to 1.5. For all calculations, we used $\phi = 5.4$ eV, and $I_{set} = 100$ pA.

To compare the model calculation with our data in Fig. S5, we extract the CDW modulation amplitude from the STM images in Fig. 2. Because the local CDW amplitude depends on the local doping and band shift, we restrict the image analysis to an area corresponding to two unit-cells approximately half way between defects A and C, where the contrast at any given bias is uniform. Meanwhile, the Fourier filtering to extract the CDW component from this small region (Fig. S5a) was calculated for the entire field of view in Fig. 2. The green shaded area highlights the low bias region between $-100$ meV and $+100$ meV where the tip-sample distance decreases dramatically in the experiments, leading to a higher probability of tip changes due to stronger tip-sample interactions. To obtain a consistent set of measurements, we intentionally did not acquire STM images at these biases. The $A_{CDW}(V)$ displayed in Fig. S5b is a subset of all the calculated amplitudes for $\Delta = 70$ meV and $\epsilon/\Delta \approx \{0, 1, 1.3\}$. The dip at negative bias in the $A_{CDW}(V)$ curves calculated for a finite gap shift $\epsilon$ corresponds to $-V_{comp}$ and marks the energy at which contrast inversion happens. We further note that for a finite $\epsilon$, there exist opposite polarity biases for which the CDW modulation amplitudes are equal. These two characteristic features of our model can be compared with experiments and provide novel clues about the CDW gap amplitude $\Delta$ and its position $\epsilon$ with respect to the Fermi level.

Note that the topographic images in Fig. 2, used to extract the experimental data points in Fig. S5, have been acquired with different current setpoints to maximize the signal to noise ratio. By analyzing additional topographic images at the same biases for different current setpoints, we verified that the amplitudes changes are minute in the range of tunneling currents considered here and that the energy at which the CI bias is observed and the behavior of $A_{CDW}$ as a function of bias voltage remain unchanged.

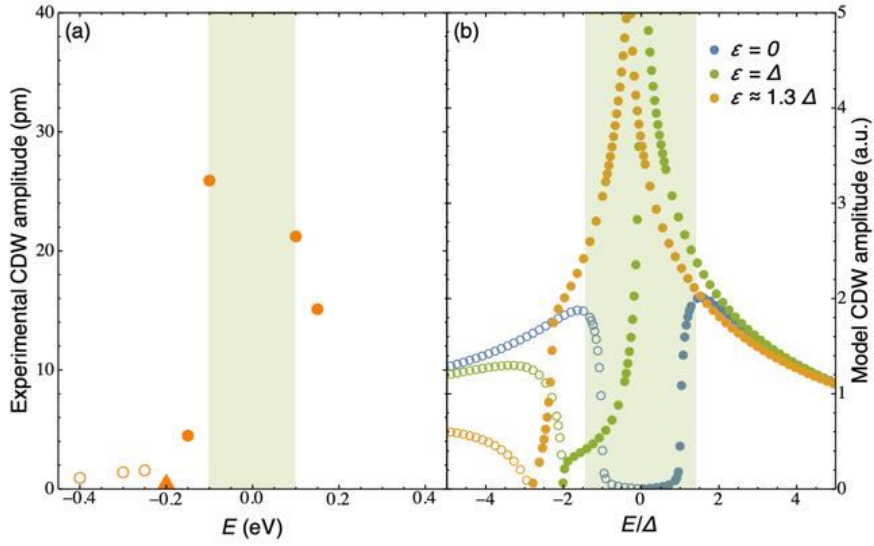

**Figure S5.** Energy dependence of the CDW amplitude. (a) Experimental CDW amplitude extracted from a two unit-cell wide region half way between defects A and B in Fig. 2. (b) CDW amplitude calculated using equation 3 for $\Delta = 70$ meV and three different shifts $\varepsilon/\Delta = \{0, 1, 1.28\}$. The best correspondence with the experimental CDW pattern and amplitudes for $\Delta = 70$ meV is achieved for $\varepsilon \cong 1.3\,\Delta$. For these values, contrast inversion is taking place at $E/\Delta \cong -2.85$. For a gap $\Delta = 70$ meV, this corresponds to $E \cong -0.2$ eV, in excellent agreement with the experimental contrast inversion around $-0.2$ eV. In both panels, the green shaded area marks the low bias region which has not been accessed experimentally, while filled (open) circles correspond to direct (inverted) real-space CDW contrast. The triangle in panel (a) marks the bias energy where no CDW contrast is observed (see Fig. 2c) corresponding to $-V_{comp}$.

The comparison of the above experimental data subset and the 1D model shows that contrast inversion takes place at $E \cong -0.2$ eV and that the same CDW patterns and comparable modulation amplitudes are measured at $\pm 100$ mV. The model best agrees with these observations for $\Delta = 70$ meV and $\epsilon/\Delta \approx 1.3$. This rough estimate is consistent with previous reports of the relative position and approximate size of the CDW gap [8, 12]. The local variation of the contrast inversion energy observed experimentally can be reproduced by tuning both the gap size and gap shift. However, a more quantitative analysis of their local variations is beyond the scope of the present study. The model that we considered here – especially in view of its simplicity – is nonetheless able to capture all experimental observations with excellent agreement. We demonstrate the ability to identify the CDW gap and its alignment with respect to $E_F$ from topographic STM images.

## VII. Large scale topography

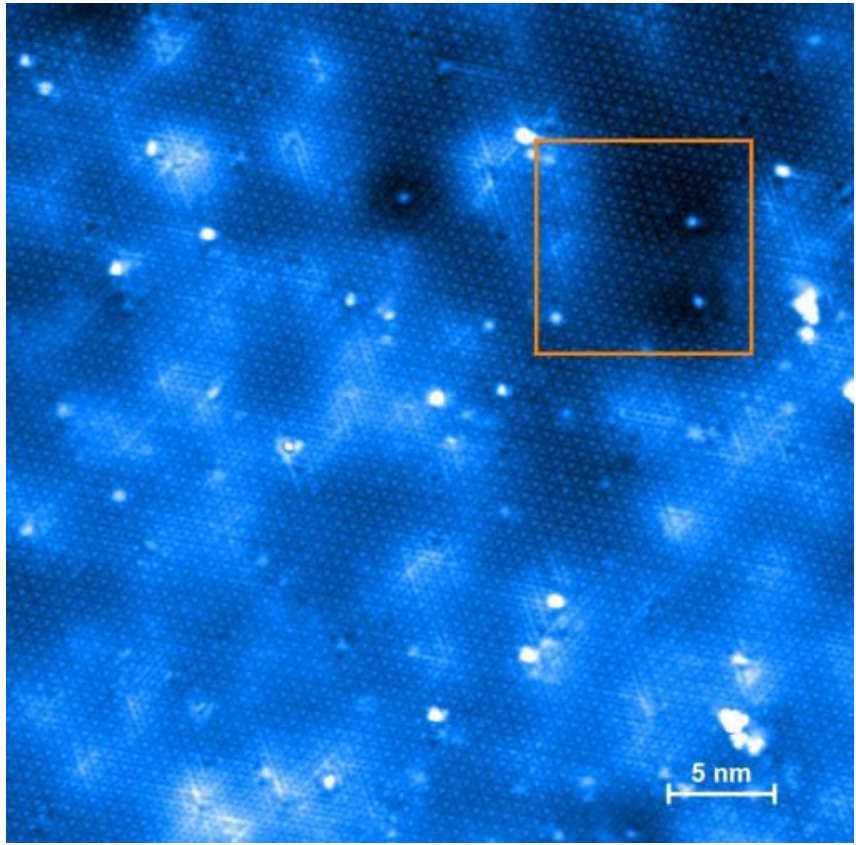

**Figure S6**. Large scale topography. The orange frame indicates the area represented in Fig. 2 of the main text. The STM tunneling conditions are very stable and no domain-wall-like phase slips are visible throughout the entire image. Scan size $40 \times 40$ nm$^2$, $V_b = 100$ mV, $I_t = 100$ pA.

# These authors made equal contributions to the work

* To whom correspondence should be addressed